# Performance Assessment of Sub-Percolating Nanobundle Network Transistors by an Analytical Model

N. Pimparkar [a]*, J. Guo [b] and M. A. Alam [a]
[a] School of Electrical and Computer Engineering, Purdue University, West Lafayette, IN 47907-1285, USA.
[b] Department of Electrical and Computer Engineering, University of Florida, Gainesville, Florida 32611-6130 USA.
*Phone: (765) 494 9034    Fax: (765) 494 6441 Email: ninad@purdue.edu

*Abstract*— Nanobundle network transistors (NBTs) have emerged as a viable, higher performance alternative to poly-silicon and organic transistors with possible applications in macroelectronic displays, chemical/biological sensors, and photovoltaics. A simple analytical model for I-V characteristics of NBTs (below the percolation limit) is proposed and validated by numerical simulation and experimental data. The physics-based predictive model provides a simple relation between transistor characteristics and design parameters which can be used for optimization of NBTs. The model provides important insights into the recent experiments on NBT characteristics and electrical purification of nanobundle networks.

## I. INTRODUCTION

Despite many recent reports of NBT whose channel material is composed of nanobundles of single wall carbon nanotubes (SWCNT)[1-6] or silicon nanowire (SiNW) in variety of applications, there is no theoretical work on the current-voltage (I-V) characteristics of these devices to allow optimization and determination of its performance limits. For NT/NW densities above the percolation limit, NBTs have emerged as a viable, higher performance alternative to poly-silicon[7] and organic transistors[8] with possible applications in macroelectronic displays, chemical/biological sensors, photovoltaics[1-6] *etc*. At densities below the percolation limit, the transistors may have applications in microelectronics and power transistors with on-current ($I_{ON}$) of the order of milliamps, much higher than a single NW/NTs devices but still with on/off ratio ~ 500 [6]. In this paper, for the first time, we propose a simple theory of I-V characteristics of NBTs (below the percolation threshold) and use the theory to model recent experiments and assess device performance.

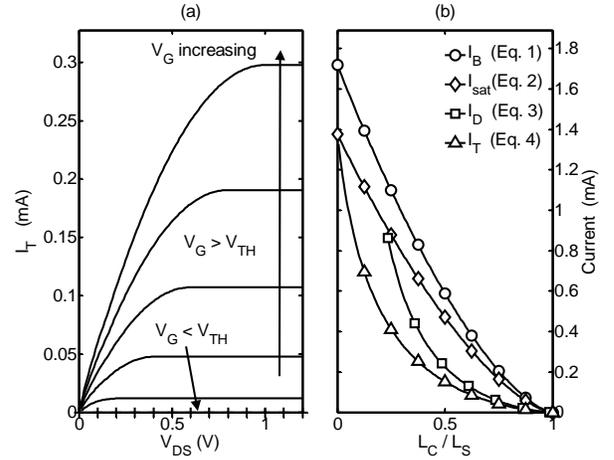

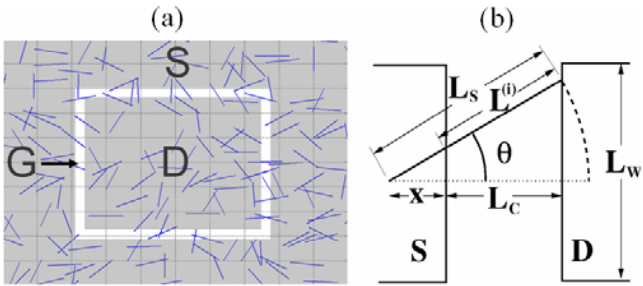

**Fig. 1**: Geometry of nanobundle network transistor: (a) A schematic of NBT. (b) A nano-stick between the S/D electrodes making an angle $\theta$ with the channel axis**.**

Fig. 2: (a) The source-drain current computed by the transmission model ($I_T$, Eq. 4) vs. $V_{DS}$ for $V_G$ - $V_{TH}$ = 0 to 1.0 $V$ with $L_C$ = 0.5 µm, $L_S$ = 1 µm, $L_W$ = 200 µm, and the stick density, $D_C$ = 1 µm$^{-2}$. (b) The on-current vs. normalized channel length computed by 4 models given by Eqs. 1-4. (i) the ballistic model, $I_B$ (Eq. 1), (ii) the velocity saturation model, $I_{sat}$ (Eq. 2), (iii) the drift-diffusion model, $I_D$ (Eq. 3), and (iv) the transmission model, $I_T$ (Eq. 4) vs. $L_C/L_S$ for $V_G$ - $V_{TH}$ = 1.0 $V$ and $V_D$ = 0.4 $V$.

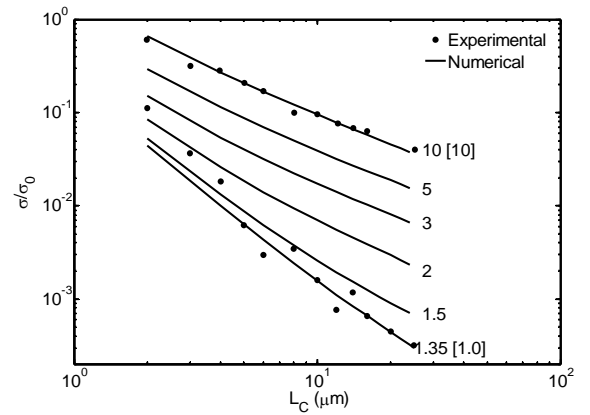

Fig. 3: Computed conductance[10] dependence on channel length for transistor above the percolation limit for different densities ($D_C$) in the strong tube-tube coupling limit is compared with experimental results[6]. The number after each curve corresponds to density of tubes, $D_C$ and the number in [ ] corresponds to $D_C$ in experiments. Parameters, chosen to reflect experimental conditions in[6] are: $L_S$ = 2µm and $L_C$ = 1–25 µm.

## II. ANALYTICAL MODEL

Consider a typical NBT[6, 9] (Fig. 1a) assembled with a bundle of nano-sticks of length $L_S$, isotropically oriented $(0 \leq \theta < 2\pi)$ onto gate oxide surface. $L_C$, $L_W$, $L^{(i)}$ and $x$ (Fig. 1b) are channel length, channel width, intercepted channel length for individual sticks and total stick overlap component with the S/D along channel axis, respectively. Assuming uniform probability of germination at all locations, the number of sticks bridging S/D and making an angle between $\theta$ and $\theta + d\theta$ with the channel axis (Fig. 1b and Fig. 6) is given by,

$$dN_S(\theta) = (D_C x) d\theta / (\pi/2) = (2D_C/\pi)(L_S \cos\theta - L_C) d\theta$$

where $D_C$ is linear tube density and $\cos\theta = (L_C + x)/L_S$. The sticks germinating at $x' > x$ distance (Fig.1b) inside the S/D boundary can not bridge S/D and make an angle of $\theta$ with channel axis, simultaneously. Hence, $(D_C x)$ is the total number of sticks germinating at a distance less than $x$ inside S/D boundary and $d\theta / (\pi/2)$ is the probability that the stick will make an angle of $\theta$ with channel axis. Total number of sticks bridging S/D, $N_S$, is given by summing over all the angles from $0$ to $\theta_{max} = \cos^{-1} L_C / L_S$. Therefore,

$$N_S(R_S) = 2D_C L_S / \pi g_B(R_S)$$

where $R_S = L_C / L_S$, and $g_B(y) = (1 - y^2)^{1/2} - y \cos^{-1} y$.

*Ballistic limit:* For NBTs in short channel limit all the sticks are in ballistic limit and the current is given by,

$$I_B = L_W C_{ox}[V_G - V_{TH}]v_T N_S \quad (1)$$

where, $v_T = (2K_B T / \pi m^*)^{1/2}$ is thermal velocity, $V_{TH}$ is threshold voltage and $C_{OX}$ is gate capacitance. The ballistic current is proportional to number of sticks, as expected.

*Velocity saturation limit:* For NBTs in high bias all the sticks can be in velocity saturation limit so that magnitude of current is independent of individual stick length. Therefore, transport is again independent of distribution of intercepted channel length and,

$$I_{sat} = L_W C_{ox}[V_G - V_{TH} - V_D/2]v_{sat} N_S \quad (2)$$

where $v_{sat}$ is saturation velocity of the carriers.

*Long channel limit:* For NBTs with longer channel limit, the transport in each of the bridging sticks is diffusive and is inversely proportional to intercepted channel length, $L^{(i)}$, for each stick. The current is given by,

$$\frac{I_D}{f(V_D,V_G)} = \int_0^{\theta_S} \frac{2D_C L_C b}{\pi} \frac{(L_S \cos\theta - L_C)}{L_C/\cos\theta} d\theta = \frac{D_C}{\pi} g_D(R_S) \quad (3)$$

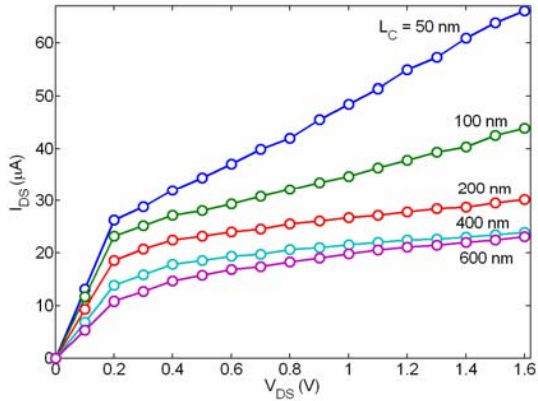

**Fig. 4:** I-V characteristics of metallic CNTs are computed by a separate Monte-Carlo simulation, which has been validated by experiments and explained in[11] in detail. For a long CNT, the current saturates at 25μA, but for a short CNT, carrier transport is quasi-ballistic and the current can be well above 25μA. This is used in Fig. 5.

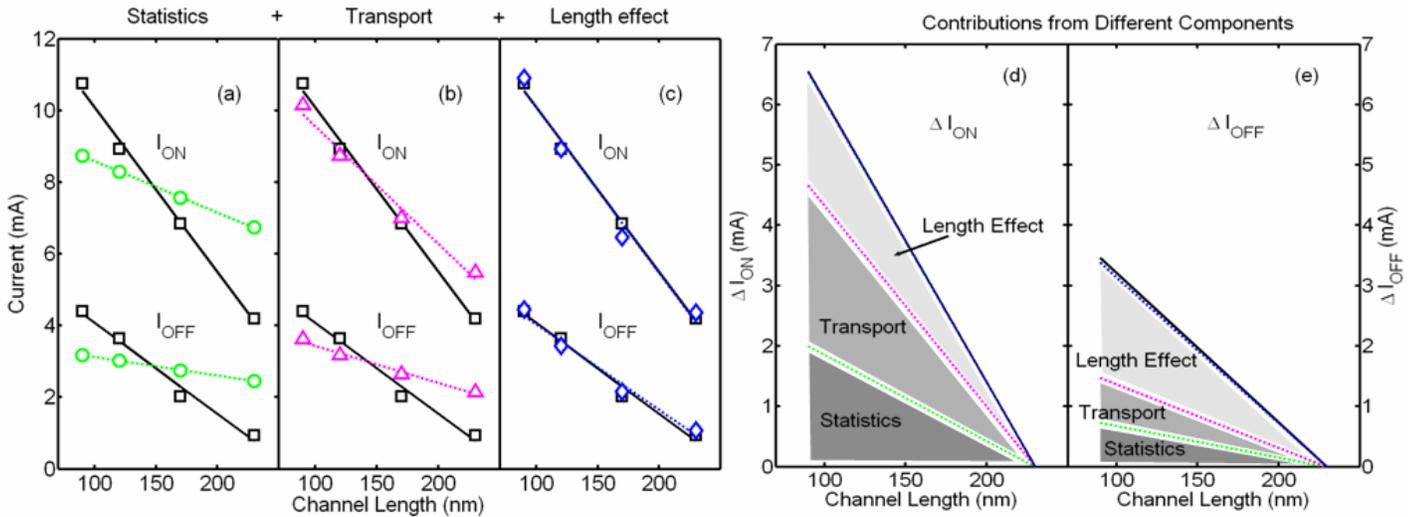

**Fig. 5:** Experimental[6] (squares with solid line fit) and theoretical (dotted line fits) $I_{ON}$ and $I_{OFF}$ vs. $L_C$. The inverse scaling of $I_{ON}$ and $I_{OFF}$ with $L_C$ is a consequence of sequential addition of (a) *Statistical effect:* Increased probability of sticks to bridge S/D as channel length decreases, (b) *Transport effect:* Prevalence of ballistic transport in shorter channel devices, and (c) *Average length effect:* Average length for m-SWCNT $<L_M>$ ~ 350 nm is made less than s-SWCNTs $<L_S>$ ~ 1000 nm. Relative contributions from the three effects in (d) $I_{ON}$ and (e) $I_{OFF}$. Here, mean free path for s-SWCNT $\lambda_D$ ~ 200 nm. Note that theory matches well with experiment only after all three effects are included in the simulation as shown in (c).

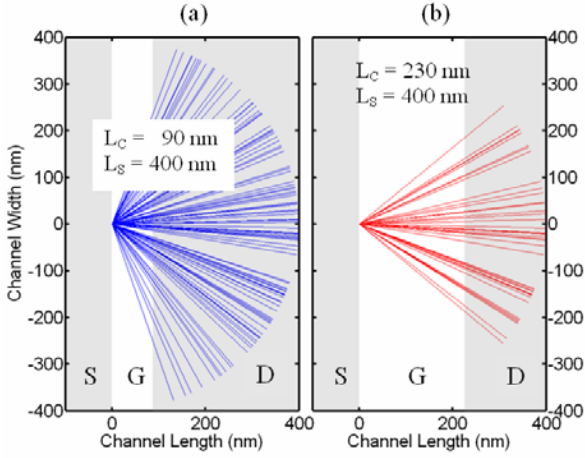

**Fig. 6.** If all sticks in Fig. 1a were collected at a point (with angles preserved), it is easy to see that *more sticks connect S/D for a transistor with a shorter channel length before the electrical burning process.* Here, stick length is $L_S = 400$ nm and channel lengths are (a) $L_C = 90$ nm and (b) $L_C = 230$ nm.

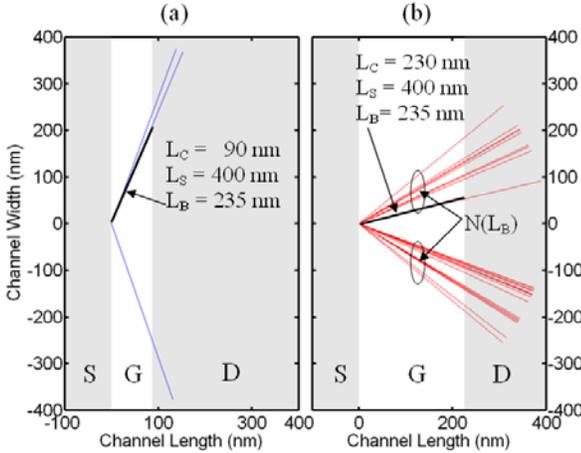

**Fig. 7.** *Less number of sticks survives the electrical burning process for a transistor with lower channel length.* Here, the burn pulse voltage is such that all the sticks with intercepted channel length, $L^{(i)}$, less than $L_B = 235$ nm are burned. The number of surviving tubes is denoted by $N(L_B)$.

where $g_D(y) = cos^{-1}y / y - (1 - y^2)^{1/2}$, $\theta_S = cos^{-1}R_S$ and $f(V_D,V_G) = \mu_0 L_W C_{OX} [(V_G-V_{TH}) V_D - V_D^2 / 2]$[10].

Finally, *for the intermediate channel lengths*, the sticks parallel to channel axis, $\theta \sim 0$, are near ballistic limit and sticks making an angle, $\theta \sim \theta_{max}$, with channel axis are near diffusive limit and the proportion of the two types of sticks changes with applied bias. Still, the I-V characteristics of any single nanostick can be described by a simple analytical expression based on a transmission point of view[11]. An exact analytical result for NBTs can be obtained by replacing $1/L^{(i)}$ with $1/(L^{(i)} + \lambda)$ where $\lambda = \min(\lambda_D, \lambda_{sat} = V_D \mu_0 / v_{sat})$ and $\lambda_D$ is the mean free path for carriers. Hence, the total current

$$\frac{I_T}{f(V_G,V_D)} = \int_0^{\theta_S} \left( \frac{2D_C}{\pi} \frac{L_S \cos\theta - L_C}{L_C/\cos\theta + L_C b} \right) d\theta \quad (4)$$

$$= \frac{2D_C}{\pi b^2} \left[ b g_B(R_S) - \cos^{-1} R_S + \frac{2(bR_S+1)}{\sqrt{b^2-1}} \tanh^{-1} \frac{(b-1)\tan(\theta_S/2)}{\sqrt{b^2-1}} \right]$$

where, $b = \lambda / L_C$ (Fig. 2a). As $\lambda \to 0$, $I_T \to I_D$ and as $\lambda \to \infty$, $I_T \to I_{sat}$. A typical transistor characteristics based on Eq. (4) is shown in Fig. 2a. Fig. 2b demonstrates that the integral model (Eq. 4) reduces to long channel limit, Eq. 3 as $L_C \to L_S$ and to velocity saturation limit, Eq. 2 as $L_C \to 0$.

### III. NUMERICAL AND EXPERIMENTAL VERIFICATION

Fig. 2a appears to be a regular MOS characteristic. One might wonder if the statistical formulation in Fig. 1 is actually necessary or would a simple model with redefined mobility will do. We performed three simulations to verify the unique predictions of the proposed model by recent experiments and to assess the performance of NBTs.

*Length Scaling:* If we generalize the model above percolation limit ($D_C > 4.236^2/\pi L_S^2$)[12], with finite tube-tube coupling, the theoretically computed channel conductivity ($\sigma$) vs. channel length ($L_C$) matches well with the experimental measurements [2, 13, 14].

$I_{ON}$ - $I_{OFF}$ *Ratio:* Inverse scaling of $I_{ON}$ and $I_{OFF}$ vs. $L_C$ in the recent measurements of CNT-NBTs below percolation limit by *Infineon* (Fig. 5) can only be explained if we account for three factors associated with random distribution of tubes:

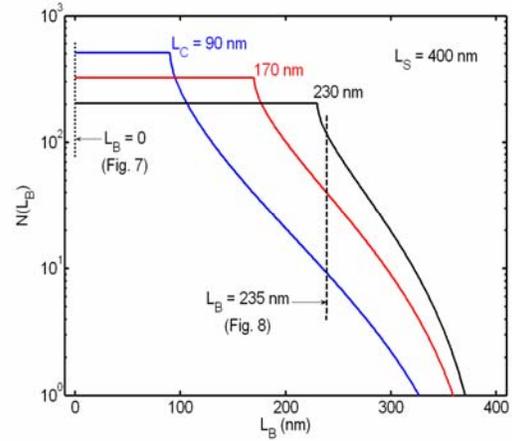

**Fig. 8.** The number of sticks, $N(L_B)$, surviving as a function of $L_B$, the maximum length of the tube burned by a burn pulse voltage of $V_B$. The definition of $N(L_B)$ is illustrated in Fig. 7b. The dotted and dashed lines correspond to the situation before (Fig. 6, $L_B = 0$) and after (Fig. 7, $L_B = 235$ nm) electrical burning, respectively. The figure clearly explains reversal of trend in on-current as a function of $L_C$ due to electrical burning (Fig. 10b).

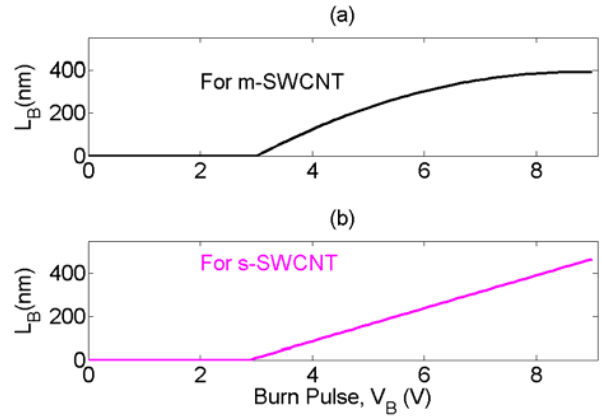

**Fig. 9**: A simple second order relation based on experimentally observed relationship[11, 15] for maximum length of the tube burned, $L_B$ vs. burn pulse voltage, $V_B$, for (a) m- and (b) s-SWCNTs when the transistor is turned off.

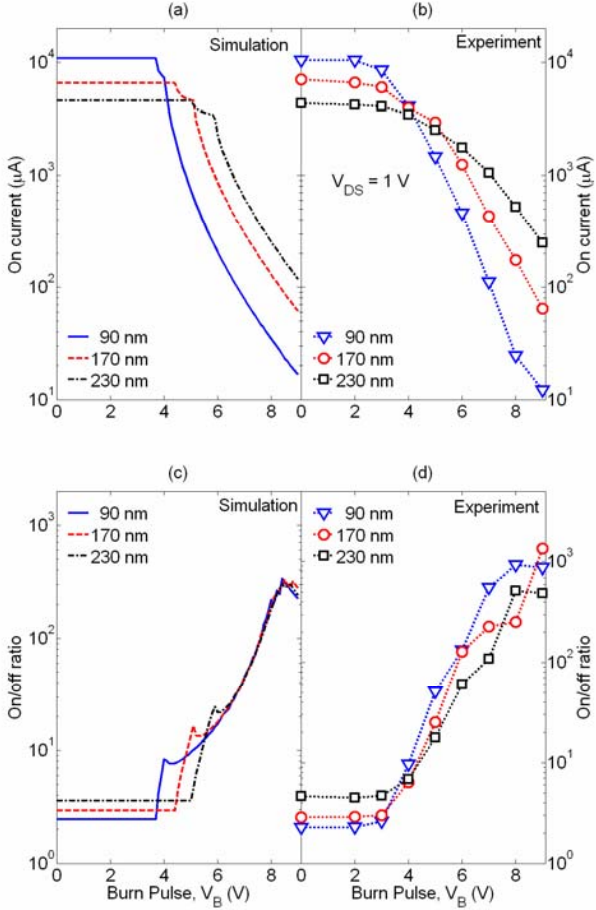

**Fig. 10**: Transistor characteristics vs. burn pulse voltage, $V_B$: Simulated (left) and experimental[6] (right) on currents (a, b) and on/off ratios (c, d) for different transistors with channel lengths ($L_C$ = 90, 170, and 230 nm) with same tube density and tube length, $L_S = 400$ nm. On current drops and on/off ratios increases with $V_B$. Although the on current scales inversely with $L_S$ before burning (Fig. 5), the trend reverses after a critical burning voltage, $V_B = 4$ V due to statistical factor (Fig. 6-8). Also the on/off ratio scales with $L_S$ which can be explained only if $<L_M> < <L_S>$. Simulation interprets experimental data quantitatively. Here, $V_{DS}$ = 1 V

(1) *Statistical effect:* Increased probability of sticks to bridge S/D as channel length decreases (Fig. 5a). More sticks bridge S/D for NBT with smaller $L_C$ (Fig. 6a) than the one with larger $L_C$ (Fig. 6b). (2) *Transport effect:* NBTs with shorter channel lengths have shorter average effective channel length giving even higher current (Fig. 5b), and (3) *Average length effect:* Fig. 10d shows that experimental on/off ratio before electrical breakdown ($V_B = 0$) is proportional to channel length. In other words, the off current ($I_{OFF}$) decreases faster than on current ($I_{ON}$) with increasing channel length. For NBTs with density lower than percolation threshold and for $L_C < L_S$, where the stick-stick interaction is not important, this effect can only be explained if average length for metallic (m)-SWCNT $<L_M>$ ~ 350 nm is less than semiconducting (s)-SWCNTs $<L_S>$ ~ 1000 nm (Fig. 5c). Fig. 5d and 5e show the relative contribution from the three effects in $I_{ON}$ and $I_{OFF}$, respectively. The total statistical effect, which is sum of (1) and (3), plays a dominant role in interpreting this experiment. Here, mean free path for s-SWCNT $\lambda_D$ ~ 200 nm. The I-V characteristics of metallic tubes are computed by a separated Monte Carlo simulation [15] and can be well described by a simple analytical expression, as shown in Fig. 4.

*Electrical Filtering of Metallic Tubes:* The model offers unique insights into and interpretation of the experiments involving widely-used electrical purification of NBTs for improving the on-off ratio[4-6, 16] (Figs. 6-10). The purification process involves removing m-SWCNTs by first applying high $V_G$ to turn-off s-SWCNTs with p-type conduction, then applying a large drain burn-pulse ($V_B$) to burn m-SWCNTs[6]. The puzzle of the experiment is that although initially (and as expected) $I_{ON}$ scales inversely with $L_C$, however beyond a threshold $V_B$, the trend reverses, and $I_{ON}$ becomes proportional to $L_C$ (Fig. 10b). Our model interprets this as a consequence of modification of the bridging probability between S/D as a function of $V_B$. The electrical burning breaks all CNTs with a length shorter than a critical burning length, $L_B$, corresponding to a given $V_B$. Given that the intercepts, $L^{(i)}$, decrease with $L_C$, more tubes burn at a given $V_B$ and the number of tubes and hence current decreases more rapidly for device with lower $L_C$. If all sticks in Fig. 1a were collected at a point (with angles preserved), it is easy to see that more sticks connect S/D for a transistor with a shorter channel length ($L_C$ = 90, 230 nm and $L_S$ = 400 nm) before the electrical burning process, Fig. 6. However, less number of sticks survives the electrical burning process for a transistor with lower channel length after a critical burning voltage, Fig. 7 ($L_B$ = 235 nm). Fig. 8 quantitatively shows the number of tubes surviving the electrical filtering for a given $L_B$. Based on experimentally observed relationship between $L_B$ and $V_B$ (Fig. 9)[15, 17, 18]**,** Fig. 10 shows that our model can quantitatively interpret measured data of $I_{ON}$ vs. $V_B$ (Fig. 10a, b) and on/off ratio, $I_{ON}/I_{OFF}$ vs. $V_B$ (Fig. 10c, d)[6]. We know of no previous interpretation of this effect in the literature.

IV. CONCLUSIONS:

We have proposed and validated a simple analytical model for I-V characteristics of NBTs. The model provides a simple relation between transistor characteristics and designable parameters, which is useful for NBT design optimization. The model also provides a theoretical framework to interpret recent experiments on transistor characteristics and electrical purifications of nanobundle networks. This could allow development of an efficient protocol for electrical purification of CNTs.

Acknowledgements:
This work was supported by the Network of Computational Nanotechnology and the Lilly Foundation. The authors would like to gratefully acknowledge discussions with S. Kumar, Prof. Murthy and Sayeed Hasan.